\documentclass[10pt,preprint2,a4paper]{aastex}
\usepackage{graphics,epsf}
\usepackage{amsmath}                
\usepackage{amsfonts}               
\usepackage{amssymb}                
\usepackage{epsfig}                 

\def \K{~\rm{K}}

\def \AU{~\rm{AU}}
\def \erg{~\rm{erg}}
\def \yrs{~\rm{yrs}}

\def \kpc{~\rm{kpc}}

\def \days{~\rm{days}}

\begin{document}

\title{AN INDICATION FOR THE BINARITY OF P CYGNI FROM ITS SEVENTEENTH CENTURY ERUPTION}

\author{Amit Kashi\altaffilmark{1}}


\altaffiltext{1}{Department of Physics, Technion $-$ Israel Institute of
Technology, Haifa 32000 Israel; kashia@physics.technion.ac.il}

\setlength{\columnsep}{1cm}

\small

\begin{abstract}
I show that the 17th century eruption of the massive luminous blue variable (LBV) star P~Cygni
can be explained by mass transfer to a B-type binary companion in an eccentric orbit,
under the assumption that the luminosity peaks occurred close to periastron passages.
The mass was accreted by the companion and liberated gravitational
energy, part of which went to an increase in luminosity.
I find that mass transfer of $\sim 0.1~\rm{M_\odot}$ to a B-type binary
companion of $\sim 3-6~\rm{M_\odot}$ can account for the energy of the
eruption, and for the decreasing time interval between the observed peaks in
the visual light curve of the eruption.
Such a companion is predicted to have an orbital period of $\sim 7$~years,
and its Doppler shift should be possible to detect with high resolution spectroscopic observations.
Explaining the eruption of P~Cygni by mass transfer further supports
the conjecture that all major LBV eruptions are triggered by interaction
of an unstable LBV with a stellar companion.
\end{abstract}

\keywords{binaries: general$-$stars: mass loss$-$stars: winds, outflows
$-$stars: individual (P Cyg)}

\section{INTRODUCTION}
\label{sec:intro}

The luminous blue variable (LBV) P~Cygni has undergone a series of eruptions
in the 17th century (de Groot 1969, 1988): a giant eruption in 1600,
followed by 4 smaller ones which started in 1655 and terminated in 1684.
Only the 1843 Great Eruption of $\eta$~Car is better documented than this eruption,
an amazing fact as the P~Cygni eruptions began before the invention of the telescope.
Being the nearest LBV, at a distance of $1.7\kpc$ (Najarro et al. 1997a),
the 3rd magnitude eruptions were seen by naked eye (de Groot 1969).

P~Cygni was commonly considered to be a single star.
Its eruption was associated with single LBV star processes
(e.g., Humphreys and Davidson 1994; Lamers \& de Groot 1992).
The peculiar morphology of the nebula which was formed by the eruption of P~Cygni
(Nota et al. 1995) lead Israelian \& de Groot (1999) to suggest that a different physical
process is responsible to the eruption of $\eta$~Car and P~Cygni.
On the other hand, Kashi et al. (2009) showed that the eruption of P~Cygni lies on a
strip in the total energy vs. timescale diagram together with other intermediate
luminosity optical transients, including the two 19th century eruptions of Eta Carinae.
This suggests that the same physical mechanism that is applicable to the eruptions of
$\eta$~Car and the other transients, accretion onto a MS companion and liberation of
gravitational energy, is responsible to the eruption of P~Cygni as well.

In Kashi \& Soker (2009) we reconstructed the evolution of $\eta$ Car in the last
two centuries, and built a model suggesting that the two 19th century eruptions were triggered
by periastron passages of the companion star.
Including mass loss by the two stars and mass transfer from the LBV to
the companion we obtained a good match of periastron passages to the two peaks
in the light curve of the $1838-1857$ Great Eruption.
Mass transfer was found as an important and crucial process in reducing
the orbital period during the eruption.
If only mass loss is included, without any mass transfer
from the erupting LBV to the companion, no match can be obtained.
Another advantage of the mass transfer model is that the
gravitational energy released by the mass accreted onto the companion
can account for the extra energy of the eruption.

Major LBV eruptions are defined as occasional very luminous eruptions at energy of
$\sim 10^{48} - 10^{50} \erg$.
The typical duration of these eruptions is $\sim 10^{2} - 10^{4} \days$.
We conjectured that all major LBV eruptions are triggered by stellar companions,
and that in extreme cases a short duration event with a huge mass transfer
rate can lead to a bright transient event on time scales of weeks to months,
(Kashi \& Soker 2009; Kashi et al. 2009).

In this paper I take the model from Kashi \& Soker (2009) and apply it to
the eruption of P~Cygni.
In section \ref{sec:model} I show that a mass transfer to a binary companion
can account for the changing period of the series of eruptions of P~Cygni,
and that the fraction of this radiation emitted in
visible wavelengths is compatible with that observed.
In section \ref{sec:discussion} I propose possible ways of detecting the companion
and discuss general implications to other LBVs.

\section{THE MODEL}
\label{sec:model}

\subsection{Historical observations and parameters}
\label{subsec:model:historical}

The model relies on historical observations of the light curve of P~Cygni from the seventeenth century.
The old light curve of P~Cygni (de Groot 1988) shows peaks in
(1) 1600.7, (2) 1654-5, (3) 1664.5, (4) 1672.6 and (5) 1679.6.
Though the light curve shows an increase in 1653.5, before peak (2),
there are evidences that this increase actually occurred in 1655 (Lamers \& de Groot 1992).
I will not refer to the eruption of peak (1),
as it seems to be a solitary eruption, not associated with the series of eruptions which followed it.
I will rather refer to peaks (2)-(5), namely to the series of smaller eruptions between 1655-1685.
The time intervals between the peaks in the series decreased with each peak and are 9.5, 8.1 and 7 $\yrs$.

Several authors tried to determine the LBV's properties.
The estimates range between $T_1=17,000-20,000\K$ for the temperature,
and $L_1 = 5.5-7.5 \times 10^5~\rm{L_\odot}$ for the luminosity.
(Lamers et al. 1983; Lamers \& de Groot 1992; Najarro et al. 1997a,b; Pauldrach \& Puls 1990).
As to the mass of the LBV, there is less agreement.
El Eid \& Hartmann (1993) found that stellar evolution considerations give $M_1=50~\rm{M_\odot}$,
while Pauldrach \& Puls (1990) found $M_1=23~\rm{M_\odot}$ from spectroscopic observations.
I will adopt the set of parameters from Najarro et al. (1997a).
The LBV radius is taken to be $R_1 \simeq 75~\rm{R_\odot}$, the luminosity
$L_1 = 5.6 \times 10^5~\rm{L_\odot}$ and the effective temperature $T_1 = 18,200\K$.

\subsection{Model assumptions and caveats}
\label{subsec:model:assumptions}

The model relies on the following assumptions:
\begin{enumerate}
\item The shortening of the time interval between the peaks is related
to interaction with a binary companion.
\item The peaks occurred at or very close to periastron passages in an eccentric orbit,
when the separation between the stars is considerably smaller than during most of the orbital period.
\item The orbital period of the companion when the eruption is terminated
was somewhat shorter then the time interval between the last two peaks.
Considering the inaccuracy of the observations and the difficulty in determining the exact time of the peaks,
I take the orbital period of the companion when the eruption is terminated to be $P_f = 7~\yrs$.
\item The periastron passage of the companion exerts tidal forces on the LBV.
The outer layers of the LBV become unstable due to internal processes unrelated to the companion.
The tidal force amplifies the instability and triggers an eruption,
causing the LBV to lose mass.
\item Part of this mass is accreted by the companion,
and liberates gravitational energy that increases the total luminosity.
In addition, the companion might blow jets that shape the nebula (Soker 2001).
\end{enumerate}

The fact that a binary companion has not been directly detected though P~Cygni has been observed for
more than 400 $\yrs$ is not difficult to explain.
A main sequence (MS) companion of $M_2 \simeq 3-6~\rm{M_\odot}$ can be easily hidden by the luminosity of the LBV,
as its luminosity would be $L_2 \simeq 100-1500~\rm{L_\odot}$, much smaller than
the luminosity of the LBV.
Such a star would have approximately the same temperature of the LBV
$T_2 \simeq 12,500-19,000 \K$ at much weaker luminosity, so it would not be easily detected
by spectroscopic or photometric observations.
As I conclude below, it might be possible to observe the companion if a continuous 7 $\yrs$ observation,
as the duration of the suggested orbital period is to be performed.

\subsection{Calculations of the orbital period change}
\label{subsec:model:calculations}

The mass lost from the system during the eruptions acts to increase the orbital period, while the
mass transferred from the LBV to the less massive companion star acts to reduce the
orbital period (Eggleton 2006).
The roles of these two competing effects were calculated for the
Great Eruption of Eta Car (Kashi \& Soker 2009).
As a result of mass transfer the following periastron passage, and therefore the next eruption,
would occur sooner than for a system with a constant orbital period.
This process repeats until the instability in the LBV stops.
From that point on the orbital period remains approximately stable,
changing only very slightly due to mass loss from the LBV and mass accretion.

As explained in section \ref{subsec:model:assumptions},
following the basic assumption that the peaks occurred at or very close to periastron passages,
I assume that the orbital period of the companion when the eruption is terminate was $P_f = 7~\yrs$.
For LBV and companion masses of $M_1 = 25~\rm{M_\odot}$ and $M_2 = 3~\rm{M_\odot}$, respectively,
the semi-major axis is $a_f = 11.1 \AU$.
The companion must pass very close to the LBV at periastron, for its gravity
to influence the LBV.
The eccentricity is limited from above by the requirement that at present even at periastron passages the
LBV does not fill its potential lobe (the analogue to a Roche lobe, but referring to an eccentric orbit).
This gives the limit $e \lesssim 0.94$.
There is no strict limit from below on the eccentricity.
However, it cannot be too small as the model requires
a clear distinction between periastron and apastron interaction,
and having potential overflow during the eruptions, when the LBV expands
to about $\sim 3$ times its radius during the eruptions.
The later requirement gives $e \gtrsim 0.88$ for our parameters,
but as the expansion factor of the LBV is not certain the lower limit for the eccentricity is not strict.
A larger expansion factor would allow a smaller lower limit.
For that I take $e=0.9$, but in any case, the results depend weakly on the exact value of $e$,
as long as it is within the range $0.88 \lesssim e \lesssim 0.94$.

During the eruption I take mass to be lost from the system, and mass transferred from the
LBV to the companion.
I follow the calculation of mass transfer and loss in a binary orbit from Eggleton (2006),
and solve the change in the orbital period back in time from the end of the eruption to the time
of its beginning.
The same calculation was performed by Kashi \& Soker (2009) to solve the orbital
parameters of $\eta$~Car during the 19th century eruptions.

The rates of change of the stellar masses (going forward in time) are
\begin{equation}
\dot M_1=-\dot m_{l1}-\dot m_t ~;~~
\dot M_2=-\dot m_{l2}+\dot m_t ~;~~
\dot M = \dot M_1 + \dot M_2,
\label{eq:Mdot}
\end{equation}
where $\dot m_{l1}$ and $\dot m_{l2}$ are the mass loss rates to infinity
from the primary and the secondary, respectively,
and $\dot m_t$ is the rate of mass transferred from the primary to the secondary.
As not much mass was lost from the system, I will assume that it was
entirely lost by the LBV, namely $\dot m_{l2}=0$.
I shall use the integral of the mass loss rates over time
\begin{equation}
M_{\rm{ej}}= \int_0^{t_{\rm{Er}}} \dot m_{l1}\,dt  ~;~~
M_{\rm{acc}} =\int_0^{t_{\rm{Er}}} \dot m_t \,dt,
\label{eq:Mejacc}
\end{equation}
were $t_{\rm{Er}}$ is the duration of the series of small eruptions.

The orbital separation is calculated as a function of time.
The orbital separation $\vec{r}$ varies according to (Eggleton 2006)
\begin{equation}
\ddot{\vec{r}}(t) = -\frac{GM(t)\vec{r}(t)}{r^3(t)} +
\dot m_t \left(\frac{1}{M_1(t)}-\frac{1}{M_2(t)} \right) \dot{\vec{r}}(t).
\label{eq:rt}
\end{equation}
The present orbital period, separation and eccentricity that I assumed serve as the initial condition.
I then perform integration backward in time to just before the series of small eruptions.
The equation cannot be solved analytically and is therefore solved numerically.
The eccentricity $e(t) \equiv |\vec{e}(t)|$ is calculated according to
\begin{equation}
GM\vec{e} =
\dot{\vec{r}}\times(\vec{r}\times\dot{\vec{r}}) -
\frac{GM\vec{r}}{r}.
\label{eq:e}
\end{equation}
The Keplerian energy per unit reduced mass $\varepsilon(t)$ is calculated according to
\begin{equation}
\varepsilon(t) = \frac{1}{2} \dot{r}^2
- \frac{GM}{r},
\label{eq:energy}
\end{equation}
and then it is possible to calculate the semi-major axis
\begin{equation}
a(t) = - \frac{GM(t)}{2\varepsilon(t)},
\label{eq:a}
\end{equation}
and the orbital period
\begin{equation}
P(t)=2\pi\sqrt{\frac{a^3(t)}{GM(t)}}.
\label{eq:P}
\end{equation}

The orbital phase at the end of the eruption, when the integration backward in time starts,
is a free parameter in the calculation.
In contrast to the case of $\eta$~Car, for P~Cygni the present date of periastron passage is unknown,
and therefore I cannot go back in time and find the orbital phase at the end of the eruption.
In my calculation I replace the free parameter of the orbital phase
at the end of the eruption with a free parameter of the periastron date at the end of the eruption
(this is an almost identical and a much cleaner way to perform the calculation).

Smith \& Hartigan (2006) found that the mass ejected in the entire eruption was $\sim 0.1~\rm{M_\odot}$.
As I deal only with the series of smaller eruptions, I will take $M_{\rm{ej}}=0.05~\rm{M_\odot}$
to be the mass lost during the eruption in my calculation.
The total bolometric energy of the entire eruption, including peak (1),
was $E_{\rm{bol,e}} \simeq 2.5 \times 10^{48} \erg$ (Humphreys et al 1999; Lamers \& de Groot 1992).
That energy is mostly radiated energy, as the kinetic energy was only $\sim 2 \times 10^{46} \erg$
(Smith \& Hartigan 2006).
Integrating the visual light curve from de Groot (1988) I get that the energy
radiated in the visual for the entire eruption is
$E_{\rm{vis,e}} \simeq 5.5 \times 10^{47} \erg \simeq 0.22 \it{E}_{\rm{bol,e}}$.
For the series of smaller eruptions only, the energy radiated in the visual is
$E_{\rm{vis,s}} \simeq 8.8 \times 10^{46} \erg$, therefore, adopting the same ratio between bolometric and visual
energy I get $E_{\rm{bol,s}} \simeq 4 \times 10^{47} \erg$.
Adding the part of the kinetic energy associated with the series of smaller eruptions
will not change this number by much.

I note that the general model, that mass transfer results in shortening of the orbital period,
does not depend on the assumption that the transferred mass is responsible for
the extra energy liberated during the eruptions.

It is very likely that only part of the energy of the accreted mass would go to
an increase in (bolometric) luminosity.
When the accreted material settles on the companion,
the gravitational energy goes to: first, warming the accreted material,
second, increasing the rotational energy of the companion, and third, inflating the companion's envelope.
I define an efficiency parameter as the ratio of the radiated
to total gravitational energy released by the accreted mass
\begin{equation}
\delta= \frac{E_{\rm{bol,s}}}{GM_2M_{\rm{acc}}/R_2}.
\label{eq:efficiency}
\end{equation}
I calibrate the mass transferred from the LBV and accreted onto the companion as
\begin{equation}
M_{\rm{acc}} \simeq 0.08 \left(\frac{\delta}{0.8}\right)^{-1}
\left(\frac {E_{\rm{bol,s}}}{4 \times 10^{47} \erg}\right)
\left(\frac{M_2}{3~\rm{M_\odot}}\right)^{-0.43}
\rm{M_\odot}.
\label{eq:Macc}
\end{equation}
where I substitute for the radius from the upper MS approximate relation of
radius to mass $R\simeq M^{0.57}$ (in solar units; Kippenhahn \& Weigert 1990).

Transferring mass from the LBV to the companion results in a decrease in the orbital period.
The ejected mass works to increase the orbital period, but for the parameters I use here
it has a much smaller effect than that of the transferred mass.
As the LBV is expected to considerably overflow its potential lobe close to periastron
(see section \ref{sec:discussion} below),
it is expected that mass accretion will take place mostly close to periastron.
It is very complicated to estimate the mass loss rate dependence on the binary
separation, as many poorly known parameters involved in this estimation.
For simplicity I assume that the mass transfer and accretion rates goes like $\sim r^{-1}$.

\begin{figure}[!t]
\resizebox{0.50\textwidth}{!}{\includegraphics{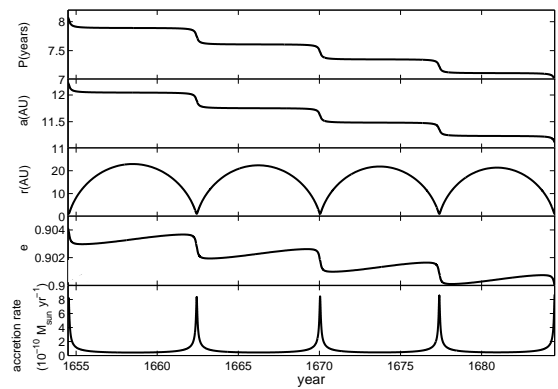}}
\caption{\footnotesize
The variation of the binary period $P$, the semi-major axis $a$, the binary separation $r$,
the eccentricity $e$, and the accretion rate (assumed to vary like $\sim r^{-1}$)
during the eruption of P~Cygni between 1654.5 and 1684.5.
I assume that the orbital period in 1684.5 was $P_f = 7 \yrs$, as the time interval between the last
two observed peaks in the series of eruptions.
The calculation involves mass transfer of $0.08~\rm{M_\odot}$ from the
$25~\rm{M_\odot}$ LBV to the $3~\rm{M_\odot}$ MS companion, and mass loss of $0.05~\rm{M_\odot}$
from the LBV to the nebula.
}
\label{fig:orbital_Eruption}
\end{figure}

\begin{figure}[!t]
\resizebox{0.50\textwidth}{!}{\includegraphics{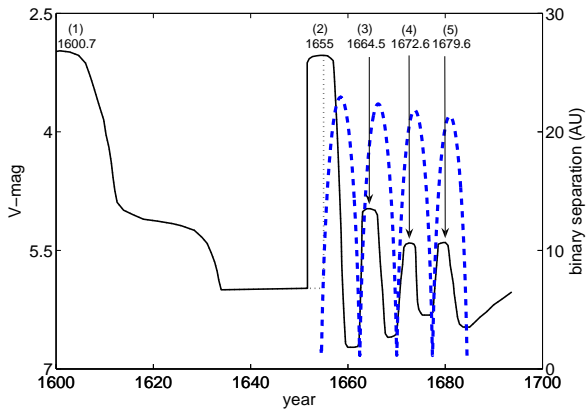}}
\caption{\footnotesize
Solid black line (left axis): the visual light curve during the 17th century eruption of P~Cygni (de Groot 1988).
The peaks in luminosity are numbered.
The time interval between the observed peaks is decreeing with time.
Though the light curve shows an increase in 1653.5, there are evidences that this increase actually
occurred in 1655 (dotted vertical line; Lamers \& de Groot 1992).
Note that the straight horizontal line between 1634 and 1652 is in fact lack of observations,
as the star was too faint during that period.
Dashed blue line (right axis): the binary separation $r$ during the eruption of P~Cygni between 1654.5 and 1684.5
(same as the third panel in Fig. \ref{fig:orbital_Eruption}).
It is assumed that orbital period of the companion when the eruption is terminated was $P_f = 7 \yrs$.
The calculation involves mass transfer of $M_{\rm{acc}} = 0.08~\rm{M_\odot}$ from the $M_1=25~\rm{M_\odot}$
LBV to the $M_2=3~\rm{M_\odot}$ MS companion (see equation (\ref{eq:Macc})), and mass loss of
$M_{\rm{ej}} = 0.05~\rm{M_\odot}$ to the nebula.
The periastron date at the end of the eruption was adjusted to 1684.5.
The efficiency factor that accounts for the fact that only part of the energy of the accreted
mass would go to an increase in (bolometric) luminosity was adjusted to $\delta=0.8$.
The mass transferred to the companion reduces the orbital period, therefore
the orbital period before the eruption is longer than at the end.
These parameters allow a fitting of a periastron passage before every peak in the light curve.
According to my model, the periastron passages of the companion exerts tidal forces on the LBV,
amplifies an inherent instability and triggers the eruptions.
The mass is accreted by the companion and liberates gravitational energy.
A fraction $\delta$ (equation (\ref{eq:efficiency})) of the gravitational energy
goes to an increase in luminosity.
As the period becomes shorter, each periastron passage occurs sooner, triggering the next
eruption in the series.
This process repeats until the LBV exits its very unstable phase.
}
\label{fig:Eruption}
\end{figure}

I find that the initial orbital period before the eruption was $P_i=8.1 \yrs$, longer than
the final orbital period $P_f$.
The variation of the binary period $P$, the semi-major axis $a$, the binary separation $r$,
the eccentricity $e$, and the accretion rate during the eruption are presented in Fig. \ref{fig:orbital_Eruption}.
In Fig. \ref{fig:Eruption} I plot together the binary separation during the eruption, and the light curve of the eruption.

I take the end of the eruption (the free parameter) to be 1684.5.
As the orbital period was continually increasing when I calculate back in time,
it does not have to reach the interval between the 1655 and 1664.5 peaks in order to
fit well to the observations, and as seen in Fig. \ref{fig:Eruption}, with the resulted $P_i=8.1 \yrs$
I get that a periastron passage occurred before every eruption, as the model requires.

\begin{figure}[!t]
\resizebox{0.50\textwidth}{!}{\includegraphics{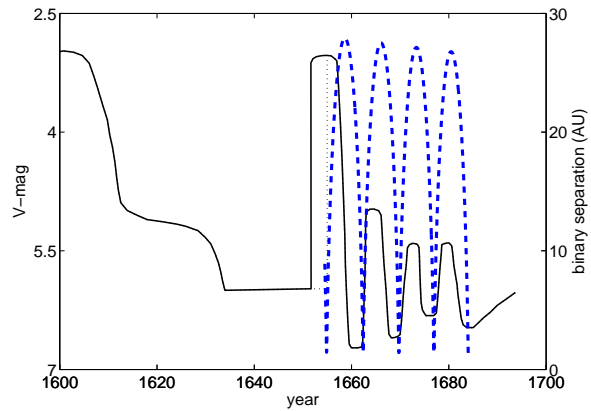}}
\caption{\footnotesize
The same as Fig. \ref{fig:Eruption}, but for other stellar masses $M_1=50~\rm{M_\odot}$ and $M_2=6~\rm{M_\odot}$
and for $M_{\rm{acc}} = 0.1\rm{M_\odot}$ transferred from the LBV to the companion.
To get a good fit, the periastron date at the end of the eruption was adjusted to 1684,
and the efficiency to $\delta= 0.5$.
The change in the masses does not at all affect the quality of the fit.
It is possible to obtain a good fit between periastron passages and eruption onsets
for this model as well.
This demonstrate the robustness of the model.
}
\label{fig:Eruption_case2}
\end{figure}

As there is uncertainty in the LBV's parameters, as well as some freedom in determining the
mass of the companion (but not the accreted mass which is obtained from the selection of the companion's mass)
I try another set of parameters.
I take the mass of the LBV to be on the upper limit $M_1=50~\rm{M_\odot}$ (El Eid \& Hartmann 1993),
and the companion mass to be $M_2=6~\rm{M_\odot}$.
A more massive companion is unlikely, as its luminosity might not be negligible compared to the LBV's,
and therefore it should have been detected by observations.
I find that a good fit to this model gives $M_{\rm{acc}} = 0.1~\rm{M_\odot}$ for the accreted mass (equation (\ref{eq:Macc})).
To obtain a good fit I slightly adjusted the periastron date at the end of the eruption, to 1684,
and used an efficiency of $\delta= 0.5$.
As illustrated in Fig. \ref{fig:Eruption_case2}, with changing the masses it is still possible to get a good fit.
Namely, it is possible to obtain periastron passage before each peak in the visual light curve.
In both cases mass transfer is a crucial ingredient of the model,
as the binary period must decrease during the eruption to obtain a good fit.

\subsection{The visible luminosity}
\label{subsec:model:visible}

According to the model, the material accreted onto the companion releases gravitational energy.
In this section I show that the fraction of this radiation emitted in
visible wavelengths is compatible with that observed.
As there are uncertainties the estimation in this section is somewhat crude.
Let us first assume that about 1 per cent of the accreted material is ejected in
disk wind from the central region of the accretion disk.
This material creates an effective larger photosphere.
The effective radius and effective temperature of the photosphere can be determined from a
set of three conjugated equations
\begin{equation}
\tau \simeq \kappa(T_{\rm{ph}},\rho_{\rm{ph}})\rho_{\rm{ph}}R_{\rm{ph}}=\frac{2}{3},
\label{eq:ph1}
\end{equation}
\begin{equation}
L_{\rm{ph}}=\frac{dE_{\rm{bol,s}}}{dt}=4\pi R_{\rm{ph}}^2 \sigma T_{\rm{ph}}^4,
\label{eq:ph2}
\end{equation}
and
\begin{equation}
\rho_{\rm{ph}}=\frac{\dot m_t}{4\pi R_{\rm{ph}}^2 v_{\rm{ph}}}.
\label{eq:ph3}
\end{equation}
where $\kappa$, $\rho_{\rm{ph}}$, $R_{\rm{ph}}$, $T_{\rm{ph}}$, $L_{\rm{ph}}$ and $v_{\rm{ph}}$ are the
opacity, density, radius, temperature, luminosity and velocity
(approximated as free-fall velocity of $1000~\rm{km~s^{-1}}$)
of the effective photosphere, respectively.
Using values for the opacity from Alexander \& Ferguson (1994),
I find an approximate solution of $T_{\rm{ph}}\simeq 10^4 \K$ and $R_{\rm{ph}}\simeq 0.3 \AU$.

The fraction of the luminosity of blackbody radiation that is observed in the
visible can be estimated by
\begin{equation}
F_{\rm{vis}} = \frac{\int_{400\rm nm}^{700\rm nm} B(T)\,d\lambda}{\int_0^\infty B(T)\,d\lambda},
\label{eq:Fvis1}
\end{equation}
where $B(T)$ is the blackbody function.
For $T_{\rm{ph}}\simeq 10^4 \K$ I get $F_{\rm{vis}} \simeq 0.33$.
In equation (\ref{eq:Fvis1}) I assumed full transmission for all visible wavelength,
and for that it is possibly somewhat overestimating the observed visible fraction.
I therefore conclude within the accuracy of the calculation and observations that
\begin{equation}
F_{\rm{vis}} \simeq E_{\rm{vis,s}}/E_{\rm{bol,s}} \simeq 0.3.
\label{eq:Fvis2}
\end{equation}
Thus, the model accounts for the visible magnitude which was observed during the eruptions.

\section{Discussion}
\label{sec:discussion}

\subsection{Mass transfer}
\label{subsec:discussion:masstransfer}

As it relies on old references, the visual magnitude light curve from the 17th century (de Groot 1988) is
not very accurate when referring to the dates where the brightening started.
For example, it is not clear when exactly the brightening of the second peak started, but
it was certainly between 1644-6 (de Groot 1969).
Nevertheless the peaks in the light curve are quite pronounced,
and the trend of decreasing time interval is very clear.
It is very unlikely that this trend is a result of observational inaccuracies.

The assumption that mass transfer and accretion onto the companion takes place close to periastron
is supported by the following argument.
Though Roche lobe is usually defined for circular orbits,
I approximate the gravitational potential at periastron as Roche potential.
The radius of the Roche lobe of the LBV is approximately (Eggleton 1983)
\begin{equation}
R_{RL1} \simeq \frac{0.49 q^{\frac{2}{3}}}{0.6 q^{\frac{2}{3}} + \ln(1+q^{\frac{1}{3}})} a
\simeq 0.73 a
\label{eq:R_RL1}
\end{equation}
where $q=M_1/M_2=25/3$ for the masses I use.
At periastron, for our parameters $R_{RL1} \simeq 0.81 \AU$
for the semi-major axis before the eruptions, and $\sim 0.73 \AU$
for the semi-major axis after the eruptions.

For present day $R_1 \simeq 75~\rm{R_\odot} = 0.35 \AU$, potential overflow does not occur , as $R_{RL1}> R_1$.
However, during the 17th century eruptions, an expansion of the LBV radius by a factor of $\sim 2.5-3$ could easily cause
the LBV to overflow its potential lobe close to periastron passages, transferring mass to the companion.
Weaker accretion probably took place through the entire event by wind accretion process (Bondi-Hoyle accretion).

The accreted mass probably had high angular momentum.
This might have led to the formation of an accretion disk and jets during the eruption of P~Cygni.
Those jets, mixed with the approximately spherical mass loss from the LBV,
may be responsible for the peculiar shape of the nebula, observed by Nota et al. (1995).
Indeed, the observations of the nebula of Smith \& Hartigan (2006) hint on some
axisymmetry, as expected from such a scenario.

\subsection{The effects of drag and tidal force}
\label{subsec:discussion:dragtidal}

In addition to mass transfer, two other effects act to reduce the orbital period,
drag force by the ejected mass that is not accreted, and tidal force by the LBV on the companion.

Drag force is exerted on the companion from the LBV ejecta that is influenced by its gravity but not accreted.
This gas resides between the accretion radius and the maximum influence radius of the companion, the cut-off radius.
Drag force cannot be the only physical process for reducing the orbital period in the case of P~Cygni for the following reasons.
(1) The mass loss rate of the LBV is relatively small
(in the Great Eruption of $\eta$~Car it was at least 200 times larger).
(2) The companion passes close to the LBV and may be prone to the drag force only close to
periastron, where it is small (since the cut-off distance is proportional to the binary separation).

Though the tide the companion exerts on the LBV is important,
the effect of the tidal force on the companion is negligible.
Soker (2005) calibrated the circularization time (Verbunt \& Phinney 1995) for
an eccentric orbit as
\begin{equation}
\begin{split}
\tau_{\rm {circ}} &= 2.5 \times 10^6
\left( \frac{f_c}{0.2} \right)^{-1}
\left( \frac{L}{10^7 \rm{L_\odot}} \right)^{-\frac{1}{3}}
\left( \frac{R}{100 \rm{R_\odot}} \right)^{\frac{2}{3}}  \\
&\times \left( \frac{M_{\rm {1,env}}}{0.01M_1} \right)^{-1}
\left( \frac{M_{\rm {1,env}}}{1 \rm{M_\odot}} \right)^{\frac{1}{3}}
\left( \frac{M_2}{0.25M_1} \right)^ {-1}  \\
&\times \left( 1+ \frac{M_2}{M_1} \right)^{-1}
\left[ \frac{a(1-e)}{3.6R} \right]^{8}
\yrs ,
\end{split}
\label{eq:taucir}
\end{equation}
where $M_{\rm {1,env}} \simeq 10 M_\odot$ is the LBV's envelope mass,
and
\begin{equation}
f_c(e) \simeq (1-e)^{\frac{3}{2}}
\left(1+\frac{15}{4}e^2 + \frac{15}{8}e^4 + \frac{5}{64}e^6 \right),
\label{eq:fce1}
\end{equation}
is a dimensionless function of the eccentricity (Hut 1982),
assuming P~Cygni rotates slowly.
For the model I suggest for P~Cygni $e=0.9$, $f_c=0.17$.
The envelope of P~Cygni is assumed to be convective.
In the unlikely case where it is radiative, then the circularization time is longer.
Using equation (\ref{eq:taucir}) I get that the circularization time for our suggested parameters, is
$\sim 2\times 10^5 \yrs$,
much longer than the duration of the eruptions.
Therefore the companion orbit is not expected to be affected by tidal force during
the eruptions.

It is expected that the P~Cygni binary system evolve into a Wolf-Rayet (WR) binary,
that would still have an eccentric orbit.
The immediate question raised is whether there exist WR binary systems with eccentric orbits,
which might have perviously resembled P~Cygni.
The VIIth catalogue of galactic WR stars (van der Hucht 2001) clearly shows that long period
WR binaries and high eccentricities correlate.
The number of very high eccentricity systems is not large, but that is expected as most systems
are short period and they have time to reduce their eccentricities during their WR stage.
Eldridge (2009) found that WR binaries having an orbital period longer than $\sim 30 \days$
are expected to remain with eccentric orbits.
This usually happens when the mass loss occurs occasionally for short periods, not having enough
time to affect the eccentricity.

The WR 140 massive binary system has an eccentricity of $e \sim 0.88$
and a period of $P \sim 7.94 \yrs$ (Marchenkoet al. 2003),
and is perhaps the most extraordinary example showing that high
eccentricity may survive the LBV stage, assuming all WR stars experience LBV evolution.

\subsection{Detecting the companion}
\label{subsec:discussion:detect}

Though the proposed model suggests interaction in a binary system,
luminous x-ray radiation is not expected.
A detailed analysis of x-ray luminosity from colliding winds (Akashi et al. 2006)
gives that a collision between the LBV and the MS companion winds, for the parameters used in the model,
is likely to produce soft x-ray at very low luminosity $< 10^{29} \erg~s^{-1}$.
The reason for that low luminosity, compared to the strong one observed in other systems
(such as $\eta$ Car; Corcoran 2005) is the very small mass loss rate of the $\sim 3-6~\rm{M_\odot}$
B-type binary companion, which makes the LBV wind dominate and consequently only very weak shocks are formed.
The companion itself might produce x-ray luminosity of $10^{27-28} \erg~s^{-1}$, unless
it is extremely magnetically active (Stelzer et al. 2005).
Bergh\"{o}fer \& Wendker (2000) analyzed ROSAT HRI observations of P~Cygni and derived an upper limit
for the flux which translates to $L_x \leq 8.4 \times 10^{30} \erg~s^{-1}$ for its x-ray luminosity,
adopting the more recent distance estimate of $1.7\kpc$ (Najarro et al. 1997a).
Therefore the x-ray luminosity expected from the presence of the companion does not
contradict observations.
A more sensitive observations might indeed detect the x-ray luminosity of $L_x \sim 10^{29} \erg~s^{-1}$
that my model predicts.

The spectra of P~Cygni in the optical and near-IR show that the
emission lines typically have broadening of $\sim 200~\rm{km~s^{-1}}$ (Najarro et al. 1997b).
It is expected in my model, that radial velocity shift in spectral lines due
to orbital motion would not be detected.
Most of the spectral lines that are observed are emitted from the LBV,
as it is much more luminous than the companion.
For the orbital parameters and stellar masses that I suggest in this paper,
the radial velocity of the LBV relative to the center of mass at periastron
is $\simeq 70~\rm{km~s^{-1}}$.
This maximal radial velocity is smaller than the broadening of the lines due to the LBV wind.
In addition, the binary system might be inclined to our line of sight, so this velocity might
be further reduced.
According to the model, presently the LBV reaches that velocity only once every $\sim 7 \yrs$,
or possibly slightly less, as there is mass transfer in the last peak,
and therefore the final orbital period should be somewhat shorter than the interval
between last two peeks.
Thus, every $\sim7 \yrs$ it should be possible to detect radial velocity changes due to
orbital motion in spectral lines, if the inclination angle is large enough.
I therefore predict that a continuous 7 year long observation of pronounced lines
may reveal a small doppler shift variation, close to the periastron passage.
It is impossible at the moment to predict the exact times of periastron passages.

The astrometric wobble should be easily detected by future telescopes.
For example, if P~Cygni is to be one of the $\sim 10^9$ galactic stars observed 70 times by
the Global Astrometric Interferometer for Astrophysics (GAIA) during its planned
5 year mission, there are favorable chances of detecting the companion.
Also, if the Space Interferometry Mission (SIM) is to be lunched and targeted to P~Cygni every
$\sim 0.5 \yrs$, it should detect the companion.
I strongly encourage to include P~Cygni in the list of observed objects for these two future missions.

\subsection{Implication to other LBVs}
\label{subsec:discussion:implication}

In Kashi \& Soker (2009) we suggested that major LBV eruptions are triggered by binary interaction.
The possible most problematic example to be considered against our claim was P~Cygni,
as it was believed to be a single star which underwent such an eruption.
In this paper I show that even the eruption of P~Cygni presents evidence of binary interaction,
and by that I strengthen the conjecture that probably all major LBV eruptions are triggered by interaction
of a stellar companion.

From the calculation in section \ref{subsec:discussion:dragtidal}, it is evident that
the circularization time for LBV systems is long, and for that it is probable that
many LBVs have binary companion in an eccentric orbit.

The parameter space of the companion stellar and orbital parameters is large, and therefore
it is expected that LBVs would have varying types of major eruptions.
However, I suggest that in most cases there will be a common characteristic, a change in the orbital period
as a result of mass transfer, that can be expressed in the light curve.

As the companion responsible for triggering the LBV eruptions,
which are mass loss episodes, it has a very
important role in the evolution of the massive LBV.
The process by which companions enhance the mass loss from LBV stars
during major eruptions accelerates the evolution of LBV stars to the WR phase.
The differences between different LBVs might be a result of differences in the binary properties,
such as orbital period, eccentricity and companion mass and wind momentum.
Particularly, the companion plays a major role in determining the destiny of the LBV and the binary system.
For example, the presence of massive circumbinary nebula might lead to a very bright supernova.
This process and its observational imprints have been theoretically studied by Kotak \& Vink (2006),
and might have been observed in SN 2006gy (Smith et al. 2007).

I thank my advisor Noam Soker for advising me in writing this paper,
and the referee John J. Eldridge for comments that helped to improve the paper.
This research was supported by the Asher Fund for Space Research at the
Technion, and the Israel Science Foundation.

\end{document}